\documentclass[aps,prd,12pt,preprint,tightenlines,superscriptaddress,
amsfonts,amssymb,amsmath,byrevtex,showpacs]{revtex4}
\begin{document}
\newcommand{\dR}{\mathbb R}
\newcommand{\dC}{\mathbb C}
\newcommand{\dS}{\mathbb S}
\newcommand{\dZ}{\mathbb Z}
\newcommand{\id}{\mathbb I}
\newcommand{\ep}{\epsilon}
\newcommand{\dV}{\mathbb V}
\newcommand{\dM}{\mathbb M}
\newcommand{\dH}{\mathbb H}

\title{Coherent state quantization of a particle \\
in de Sitter space}

\author{Jean-Pierre Gazeau }
\affiliation{LPTMC and F\'ed\'eration de Recherches APC, Boite 7020,
Universit\'{e} Paris 7 Denis-Diderot, 75251 Paris Cedex 05,
France; e-mail: gazeau@ccr.jussieu.fr}
\author{W\l odzimierz Piechocki}
\affiliation{So\l tan Institute for Nuclear Studies, Ho\.{z}a 69,
00-681 Warszawa, Poland; e-mail: piech@fuw.edu.pl}

\date{\today}

\begin{abstract}
We present a coherent state quantization of the dynamics of a
relativistic test particle on one-sheet hyperboloid embedded in
three-dimensional Minkowski space. The group $SO_0(1,2)$ is
considered to be the symmetry group of the system. Our procedure
relies on the choice of  coherent states of the motion  on a
circle. The coherent state realization of the principal series
representation of $SO_0(1,2)$ seems to be a new result.
\end{abstract}
\maketitle

\section{Introduction}

In this paper we carry out a coherent state quantization of the
dynamics of a relativistic test  particle  on a one-sheet
hyperboloid embedded in three-dimensional Minkowski space, i.e. on
the two-dimensional de Sitter space with the topology
$\dR\times\dS $.

Quantizing the dynamics of a free particle in a curved spacetime
is not a purely academic issue since it occurs in fundamental
problems. For instance, there is a great  interest in cosmology
owing to the puzzle of dark matter and energy (see \cite{WMAP,MST}
and references therein). Models constructed to solve it within the
framework of higher dimensional theories assume that baryonic
matter occurs only on one brane embedded in a higher dimensional
space (see, e.g. \cite{DG,D1,D2,LRS}). It is important in the
context of  brane cosmology to understand the confinement problem
of a particle to the brane. Examination of the classical particle
confinement \cite{SSS,HL} is not sufficient because elementary
particles are quantum objects. Quantization of particle dynamics
on two-dimensional hyperboloid embedded in three-dimensional
Minkowski space may be treated to some extent as a toy model of
this problem. Another example is quantization of particle dynamics
in singular spacetimes in order to see what one can do to avoid
problems connected with singularities in quantum theory. The case
of removable-type singularities has been considered in
\cite{WP2,WP}. It is important to extend these analysis to
spacetimes with essential-type singularities because the results
may bring some new ideas useful in the construction of quantum
gravity.

The problem of a particle in de Sitter space has already been
solved rigorously within the group theory oriented quantization
scheme \cite{WP,WP1}. Since the coherent state method seems to be
powerful but it is still under development, it makes sense to
compare its effectiveness with the group theoretical one. It is
the goal of this paper. In what follows we use the results of
\cite{WP,WP1} concerning the classical dynamics of a particle. The
quantization method described in \cite{JP4,JP5} is applied to find
the corresponding quantum dynamics. More specifically, we test a
method based on adapted coherent states and inspired by the
Berezin approach \cite{JP1,AMP,JP2,JPG,JP3}. It seems to be
applicable to situations when the canonical quantization method
fails owing to, e.g., the  operator ordering (see \cite{KFT} and
references therein) and irreducibility of representation problems
\cite{PW}.

Our paper is organized as follows: In section II we specify the
canonical structure of our system which is suitable for
quantization in the way `first reduce and then quantize'.  The
choice of coherent states of a particle on hyperboloid, based on
the choice of coherent states of a particle on a circle, is
presented in section III. Both the coherent states and quantum
observables are parametrized by some real parameter $\ep > 0$. We
show in section IV that in the limit $\ep \rightarrow 0$ our
coherent state method leads to the Lie algebra homomorphism of
classical observables into quantum observables. In section V we
relate our representation to the Bargmann principal series
representation of $SU(1,1)$ group. We shortly discuss our results
in section VI. The two appendices should help the reader to follow
easily some technical aspects of the paper.

\section{Phase space and observables}

In the context of this paper the coherent state quantization will
play the role of a canonical quantization. Thus, it needs the
specification of canonical phase space, observables and symmetries
of the system. To make the paper self-contained, we recall the
main steps of our paper \cite{WP} concerning the classical
dynamics:

The two-dimensional de Sitter space $\dV$ with the topology
$\dR\times\dS$ may be visualized as a one-sheet hyperboloid
$\dH_{r_0} $ embedded in 3-dimensional Minkowski space $\dM $,
i.e.
\begin{equation}\label{hyp}
\dH_{r_0}:=\{(y^0,y^1,y^2)\in \dM~|~ (y^2)^2+(y^1)^2 - (y^0)^2
=r_0^2,~r_0 >0\},
\end{equation}
where $r_0$ is the parameter of the one-sheet hyperboloid
$\dH_{r_0} $.

\noindent It is commonly known that the induced metric,
$g_{\mu\nu}~(\mu ,\nu =0,1)$, on $\dH_{r_0} $ is the de Sitter
metric.

An action integral, $\mathcal{A}$, describing a free relativistic
particle of mass $~m_0 >0$  in gravitational field  $g_{\mu \nu}$
is proportional to the length of a particle world-line and is
given by
\begin{equation}\label{lag}
\mathcal{A}=\int_{\tau_1}^{\tau_2}~L(\tau)~d\tau,~~~~~L(\tau)
:=-m_0 \sqrt{g_{\mu\nu}\dot{x}^\mu \dot{x}^\nu} ,
\end{equation}
where $\tau$ is an evolution parameter, $x^\mu$ are intrinsic
coordinates and   $\dot{x}^\mu := dx^\mu/d\tau $. It is assumed
that $ \dot{x}^0>0 $, i.e., $x^0$ has interpretation of time
monotonically increasing with $\tau$.

The action (2) is invariant under the reparametrization
$ \tau\rightarrow f(\tau)$ of the world-line (where $f$ is an arbitrary
function of $\tau$).
This gauge symmetry leads to the constraint
\begin{equation}\label{con1}
    G:= g^{\mu\nu}p_\mu p_\nu - m_0^2=0,
\end{equation}
where $g^{\mu\nu}$ is the inverse of $g_{\mu\nu}$ and $p_\alpha :=
\partial L/\partial\dot{x}^\alpha ~ (\alpha =\mu ,\nu )$ are
canonical momenta.

Since a test particle does not modify spacetime, the local
symmetry of spacetime coincides with the algebra of all Killing
vector fields.  It is known (see, e.g. \cite{SH}) that a Killing
vector field $Y$ may be used to find a dynamical integral $D$ of a
test particle moving along a geodesic by
\begin{equation}\label{dyn}
D=p_\mu Y^\mu,~~~~~\mu= 0,1 ,
\end{equation}
where $Y^\mu$ are components of $Y$.

To be more specific we parametrize the hyperboloid (\ref{hyp}) as
follows \cite{WP}
\begin{equation}\label{par}
y^0=-\frac{r_0\cos \rho /r_0}{\sin \rho
/r_0},~~~~y^1=\frac{r_0\cos \theta /r_0} {\sin \rho
/r_0},~~~~y^2=\frac{r_0\sin \theta /r_0}{\sin \rho /r_0},
\end{equation}
where $0<\rho < \pi r_0 $ and $0\leq \theta < 2\pi r_0$.

For the parametrization (\ref{par}) the metric tensor induced on
$\dH_{r_0}$ reads
\begin{equation}\label{met}
ds^2 = (d\rho^2 - d\theta^2)\sin^{-2}(\rho/r_0).
\end{equation}
Thus the constraint (\ref{con1}) has the form
\begin{equation}\label{con2}
G= (p_\rho^2 - p_\theta^2)\sin^2(\rho/r_0)-m_0^2 =0,
\end{equation}
where $p_\rho :=\partial L/\partial\dot{\rho}$ and $p_\theta
:=\partial L/\partial\dot{\theta}$ are canonical momenta.

The Killing vector fields $Y_a ~(a=0,1,2)$ of our two-dimensional
de Sitter space may be found \cite{WP}  by specification of the
infinitesimal transformations of the proper orthochronous Lorentz
group $SO_0(1,2)$. These transformations, in parametrization
(\ref{par}), read
\begin{equation}\label{t1}
(\rho,~\theta)\longrightarrow(\rho,~\theta+a_0 r_0),
\end{equation}
\begin{equation}\label{t2}
(\rho,~\theta)\longrightarrow(\rho-a_1 r_0 \sin \rho /r_0~\sin
\theta /r_0,~\theta+a_1 r_0\cos \rho /r_0~\cos \theta /r_0),
\end{equation}
\begin{equation}\label{t3}
(\rho,~\theta)\longrightarrow(\rho+a_2 r_0\sin \rho /r_0~\cos
\theta /r_0,~\theta+a_2 r_0\cos \rho /r_0~\sin \theta /r_0),
\end{equation}
where $(a_0,a_1, a_2) \in \dR^3 $ are small parameters. The
transformation (\ref{t1}) corresponds to the infinitesimal
rotation interpreted here as space de Sitter translation, whereas
(\ref{t2}) and (\ref{t3}) define  two infinitesimal boosts. One of
them can be interpreted as Lorentz boost, another one describes
   de Sitter `time' translation.

The dynamical integrals $J_a ~(a=0,1,2)$ defined by (\ref{dyn})
and corresponding to (\ref{t1})-(\ref{t3}), respectively,  read
\begin{equation}\label{d0}
J_0=p_{\theta}~r_0,
\end{equation}
\begin{equation}\label{d1}
J_1=-p_\rho ~r_0\sin \rho /r_0~\sin \theta /r_0 + p_\theta
~r_0\cos \rho /r_0~\cos \theta /r_0,
\end{equation}
\begin{equation}\label{d2}
J_2= p_\rho ~r_0\sin \rho /r_0~\cos \theta /r_0 + p_\theta
~r_0\cos \rho /r_0~\sin \theta /r_0.
\end{equation}

\noindent Making use of  (\ref{d0})-(\ref{d2}) one may rewrite the
constraint (\ref{con2}) as
\begin{equation}\label{con3}
J_2^2 + J_1^2-J_0^2  - \kappa^2 = 0,~~~~~~\kappa :=m_0 r_0 .
\end{equation}

One can verify  \cite{WP} that Eqs.\:(\ref{par}) and
(\ref{d0})-(\ref{d2}) lead to the algebraic equations
\begin{equation}\label{tra}
J_a y^a =0,~~~~~J_2 y^1 - J_1 y^2 =r_0^2 p_\rho
\end{equation}
which determine a particle geodesic for  given values of
$J_a~(a=0,1,2)$. Equations (15) do not `underdetermine' a particle
geodesic because the equation
\begin{equation}\label{hyp1}
(y^2)^2+(y^1)^2 - (y^0)^2 =r_0^2
\end{equation}
defining the spacetime $\dH_{r_0} $ should be satisfied too.

The physical phase space $\Gamma$ is defined in \cite{WP} to be
the space of all particle geodesics consistent with the constraint
(\ref{con1}). Since each triple $(J_0,J_1,J_2)$ satisfying
(\ref{con3}) defines uniquely a particle geodesic, by solution of
(\ref{tra}), the one-sheet hyperboloid (\ref{con3}) represents
$\Gamma$.  For the same reason it is natural to choose $J_a ~
(a=0,1,2) $ to represent the basic observables of our system. One
may further argue that we need to make the measurement of the
numerical values of all three observables to identify the geodesic
of a particle.

It is worthy to compare the present situation with its Minkowskian
counterpart. In the latter case there are also three Killing
vector fields and corresponding dynamical integrals: $P_0$ (time
translation), $P$ (space translation) and $K$ (Lorentz boost). Due
to the Minkowskian space homogeneity the Lorentz boost is free of
constraints  so we are left with mass-shell condition  which reads
$~P_0^2 -P^2 - m_0^2 = 0~$  and is the Minkowskian counterpart of
(\ref{con3}).

The system of a free particle in curved spacetime defined by the
action (\ref{lag}) may be treated as a gauge system in which the
gauge invariance is the reparametrization invariance. It is a
characteristic feature of such a system that the Hamiltonian
corresponding to the Lagrangian (\ref{lag}) identically vanishes.
   The general treatment of such gauge system within the constrained
Hamiltonian formalism and the reduction scheme to gauge invariant
variables has been presented elsewhere (see \cite{GJWP} and
references therein). Here we only outline the method leading to
the phase space with independent canonical variables:

The Hamiltonian formulation of a theory with gauge invariant
Lagrangian \cite{LDF,MHT} leads to an extended $2N$ dimensional
phase space $\Gamma_e$ and $M$ first-class constraints ($N$
denotes spacetime dimension and equals $2$ in our case, the
constraint is defined by (\ref{con1}) so $M=1$).  The constraint
surface $\Gamma_c$, defined e.g. by (\ref{con2}) in $\Gamma_e$
parametrized by $(\rho , \theta ,p_\rho, p_\theta)$, plays special
role in the formalism. Our type of system may have up to $2N-2M$
(that equals $2$ in our case) gauge invariant functionally
independent variables on $\Gamma_c$  which may be used to
parametrize the physical phase space and gauge invariant
observables \cite{LDF}. It is known \cite{GJWP} that the dynamical
integrals may be used to represent such variables. The observables
$J_0 , J_1 $ and $J_2$ are gauge invariant (each of them has
vanishing Poisson bracket with $G$ on $\Gamma_c$) and any two of
them are functionally independent on $\Gamma_c$ due to
(\ref{con3}). However, such two variables do not have the
canonical form (i.e., they do not form a conjugate pair) used in a
group theoretical quantization scheme.   There exists a general
method of finding the corresponding canonical variables, but it is
quite involved \cite{GJWP}. In what follows we recall the simple
method used in \cite{WP}. It consists of three steps:

\noindent First, we identify the algebra the basic observables
$J_a~(a=0,1,2)$ satisfy  on $\Gamma_c$. Direct calculations lead
to
\begin{equation}\label{alg}
\{J_0,J_1\}=-J_2,~~~\{J_0,J_2\}=J_1,~~~\{J_1,J_2\}=J_0 ,
\end{equation}
which means that our basic observables satisfy $sl(2,\dR)$
algebra.

\noindent Second, we find that the physical phase space $\Gamma$,
identified with the hyperboloid (\ref{con3}), is diffeomorphic to
the manifold $X$ defined to be
\begin{equation}\label{sp}
X:=\{x \equiv (J,\beta)~|~J \in \dR, ~ 0\leq \beta <2\pi\},
\end{equation}
where the diffeomorphism  is given by
\begin{equation}\label{dyn2}
J_0 := J,~~~J_1:=J\cos\beta - \kappa\sin\beta,~~~J_2:=J\sin\beta +
\kappa\cos\beta,~~~0<\kappa <\infty .
\end{equation}
Third, we find that the Poisson bracket on $\Gamma_c$ in terms of
the variables $J$ and  $\beta$ reads
\begin{equation}\label{po}
\{\cdot,\cdot\}:=\frac{\partial\cdot}{\partial
J}\frac{\partial\cdot} {\partial
\beta}-\frac{\partial\cdot}{\partial
\beta}\frac{\partial\cdot}{\partial J}~.
\end{equation}
Since $\{J,\beta\}=1$, the variables $J$ and $\beta$ are canonical
and $X$ will be called a canonical phase space in what follows.
The specification of the canonical structure of our system is now
complete.

The coherent state quantization does not require the independent
degrees of freedom to form  conjugate pairs. But we are going to
use the same canonical structure as in the case of group
theoretical quantization to make straightforward the comparison of
the results obtained by both methods.

Now, let us identify the symmetry group of the system. The local
symmetry  is defined by the algebra (\ref{alg}), but it is unclear
what the global symmetry  could be because there are infinitely
many Lie groups having $sl(2,\dR)$ as their Lie algebras. The
common examples are: $SO_0(1,2)$, the proper orthochronous Lorentz
group; $SU(1,1)\sim SL(2,\dR)$, the two-fold covering of
$SO_0(1,2)\sim SU(1,1)/\dZ_2$; $\widetilde{SL(2,\dR)}$, the
infinite-fold covering of $SO_0(1,2)\sim \widetilde{SL(2,\dR)}/\dZ
$ (the universal covering group).

Assuming that the symmetry of the system of a particle in de
Sitter space is defined by continuous transformations only, it is
difficult to identify the symmetry group. However, we are free to
use discrete symmetries of our system as well. Since the system of
a particle on a hyperboloid is a non-dissipative one, it must be
invariant with respect to time-reversal transformations. We have
shown in \cite{WP1} that making use of this symmetry leads to the
conclusion that the global symmetry of our system may be either
the group $SO_0(1,2)$ or $SU(1,1)$.

For the purpose of clarity of presentation of our coherent state
results, we carry out further discussion assuming that the
symmetry group is  $SO_0(1,2)$. The case of $SU(1,1)$ symmetry
will be considered elsewhere \cite{JPW}.

The problem of quantization of the basic observables
(\ref{d0})-(\ref{d2}) reduces to the problem of finding an
(essentially) self-adjoint representation of $sl(2,\dR)$ algebra
integrable to an irreducible unitary  representation of
$SO_0(1,2)$ group.

\section{Choice of coherent states}

In this section we introduce the quantities which are used in the
coherent state quantization \cite{JP4,JP5}. First, we define a
measure on $X$ which depends on some non-negative real parameter
$\ep$ as follows
\begin{equation}
\mu_{\ep} (dx):= \sqrt{\frac{\ep}{\pi}}\frac{1}{2\pi}e^{-\ep J^2}d\beta
dJ.
\end{equation}
Next, we introduce an abstract separable Hilbert space $\mathcal{H}$
with an orthonormal basis \\ $\{|m>\}_{m\in \dZ} $,  i.e.
\begin{equation}
<m_1|m_2>= \delta_{m_1,m_2},~~~\sum_{m=-\infty}^{+\infty}|m><m|=\id .
\end{equation}
Then, we construct an orthonormal set of vectors
$\{\phi_m^\ep\}_{m\in \dZ}$ which spans a Hilbert subspace
$\mathcal{K}_\ep\subset L^2(X,\mu_\ep)$, which is peculiar to our
physical system.

Being inspired by the choice of the coherent states for the motion
of a particle on a circle \cite{DBG,BRS,YSK,KKJ,JAG}, we define
$\phi_m^\ep $ by
\begin{equation}
\phi_m^\ep(J,\beta):= \exp(-\frac{\ep m^2}{2})\exp(\ep mJ-im\beta).
\end{equation}
One may check that
\begin{align}\label{theta}
\nonumber \mathcal{N}_\ep(x)&\equiv\mathcal{N}_\ep(J,\beta):=
\sum_{m=-\infty}^{+\infty}|\phi_m^\ep |^2 =\sum_{m=-\infty}^{+\infty}
\exp\left(\ep(2mJ-m^2)\right)\\
&=\vartheta_3 (i\ep J, e^{-\ep})< \infty ,
\end{align}
where the elliptic theta function \cite{MOS} is given by
\begin{equation*}
\label{ }
\vartheta_3(z,q) = \sum_{m = -\infty}^{ +\infty} q^{m^2} e^{2 m i z} .
\end{equation*}

Finally, we define the coherent states $|x,\ep>$ as follows
\begin{equation}
X\ni x \longrightarrow |x,\ep>\equiv |J,\beta,\ep>:=
\frac{1}{\sqrt{\mathcal{N}_\ep
(J,\beta)}}\sum_{m=-\infty}^{+\infty}\phi_m^\ep (J,\beta)|m>\:\in
\mathcal{H}_\ep,
\end{equation}
where $\mathcal{H}_\ep $ is an abstract Hilbert space spanned by
$\{|x,\ep>\}_{x\in X}$.

\noindent The states defined by (25) are easily verified to be
\emph{coherent}
     in the sense that they satisfy the normalization condition
\begin{equation}
<x,\ep|x,\ep>=\frac{1}{\mathcal{N}_\ep}\sum_{m_1, m_2}<m_1|
\overline{\phi_{m_1}^\ep} \phi_{m_2}^\ep|m_2>=1 ,
\end{equation}
and lead to the resolution of the identity in $\mathcal{H}_\ep$
\begin{equation}
\int_X \mu_\ep(dx)\mathcal{N}_\ep(x)|x,\ep><x,\ep| =
\sqrt{\frac{\ep}{\pi}}
\frac{1}{2\pi}\int_{0}^{2\pi}d\beta\int_{-\infty}^{\infty}dJ
\mathcal{N}_\ep(J,\beta) e^{-\ep J^2}
     |J,\beta,\ep><J,\beta,\ep|=\id .
\end{equation}
Localization properties of our coherent states are well
illustrated by the behavior of their respective `overlaps' $\:
<J',\beta',\ep|J,\beta,\ep>\:$  as functions of $(J', \beta')$
\begin{equation}
\label{overlap} <J',\beta',\ep|J,\beta,\ep> \equiv \Psi^\ep_{J,
\beta}(J', \beta') = \frac{\vartheta_3 (i\ep \frac{J + J'}{2} +
\frac{\beta - \beta'}{2}, e^{-\ep})}{\sqrt{\vartheta_3 (i\ep J,
e^{-\ep})\, \vartheta_3 (i\ep J', e^{-\ep})}}.
\end{equation}
A comprehensive discussion of this issue  will be presented
elsewhere \cite{JPW}.
\section{Homomorphism}

In the coherent state quantization method \cite{JP4,JP5} a quantum
operator is defined by the mapping
\begin{equation}
\label{oper}
   f \longrightarrow  \hat{f}^\ep :=A_\ep(f):=
\int_X \mu_\ep(dx)\mathcal{N}_\ep (x)f(x)|x,\ep><x,\ep| ,
\end{equation}
where $f = f(J, \beta)$ is a classical observable, and where the
integral is interpreted in the weak sense whenever it exists. In
general, an arbitrary unbounded operator on the Hilbert space
$\mathcal{H}_\ep $ may or may not possess a pseudo-diagonal
representation like (\ref{oper}). The domain of the operator
$\hat{f}^\ep$ is defined to be a dense subspace of
$\mathcal{H}_\ep $. It is clear that an explicit form of this
domain depends on the properties of the function $f$ (
integrability, $C^{\infty}$, etc ).

Making use of the formulae of  appendix A we obtain
\begin{equation}
\hat{J}_0^\ep =A_\ep (J_0)= \sum_{m=-\infty}^{+\infty} m |m><m|,
\end{equation}
\begin{equation}
\hat{J}_1^\ep = A_\ep (J_1)=\frac{1}{2}e^{-\ep
/4}\sum_{m=-\infty}^{+\infty}
\left((m+\frac{1}{2}+i\kappa)|m+1><m| + c. c. \right) ,
\end{equation}
\begin{equation}
\hat{J}_2^\ep = A_\ep (J_2)=
\frac{1}{2i}e^{-\ep/4}\sum_{m=-\infty}^{+\infty}
     \left((m+\frac{1}{2}+i\kappa)|m+1><m| - c. c. \right) ,
\end{equation}
where $c. c. $ stands for the complex conjugate of the preceding term.

The commutation relations corresponding to (17) are found to be
\begin{equation}
\label{qcom1}
[\hat{J}_0^\ep,\hat{J}_1^\ep]=i\hat{J}_2^\ep=-iA_\ep (\{J_0,J_1\}),
\end{equation}
\begin{equation}
\label{qcom2}
[\hat{J}_0^\ep,\hat{J}_2^\ep]=-i\hat{J}_1^\ep=-iA_\ep (\{J_0,J_2\}),
\end{equation}
\begin{equation}
\label{qcom3}
[\hat{J}_1^\ep,\hat{J}_2^\ep]=-i e^{-\ep /2}\hat{J}_0^\ep =e^{-\ep /2}
\left( -iA_\ep (\{J_1,J_2\})\right).
\end{equation}
Now, we consider the asymptotic case   $\ep \rightarrow 0$. All
operators and equations  in this limit are defined in the abstract
Hilbert space $\mathcal{H}$. The equations (\ref{qcom1})-(\ref{qcom3})
prove that in
the asymptotic case the mapping (\ref{oper}) is a homomorphism.

\section{Representation}

We prove in appendix B that our representation of $sl(2,\dR)$
algebra, at the limit $\ep \rightarrow 0$, is essentially
self-adjoint.

The classification of all irreducible unitary representations of
$SO_0(1,2)$ group has been done by Bargmann \cite{VB}. To identify
our representation within Bargmann's classification, we consider
the Casimir operator. The classical Casimir operator $C$ for the
algebra (17) has the form
\begin{equation}
C=J_2^2 +J_1^2 -J_0^2 =\kappa ^2 .
\end{equation}
Making use of (\ref{oper}) we map $C$ into the corresponding
quantum operator $\hat{C}$ as follows
\begin{eqnarray}
\hat{C}= \lim_{\ep\rightarrow 0}\hat{C}^\ep&:=&\lim_{\ep\rightarrow 0}
\left(\hat{J}_2^\ep \hat{J}_2^\ep +\hat{J}_1^\ep \hat{J}_1^\ep -
\hat{J}_0^\ep \hat{J}_0^\ep \right)= \nonumber \\
& &\lim_{\ep\rightarrow 0}\sum_{m=-\infty}^{+\infty} \left(e^{-\ep /2}
(m^2 +\kappa ^2 +
\frac{1}{4})- m^2 \right)|m><m|=\nonumber \\& & (\kappa ^2 +\frac{1}{4})
\sum_{m=-\infty}^{+\infty} |m><m| =(\kappa ^2 + \frac{1}{4})\id =:q\id .
\end{eqnarray}
Thus, we have obtained that our choice of coherent states (25) and
the mapping (\ref{oper}) lead, as $\ep \rightarrow 0$, to the
representation of $sl(2,\dR)$ algebra with the Casimir operator to
be an identity in $\mathcal{H}$ multiplied by a real constant $1/4
< q < \infty$.

In the asymptotic case,
$\hat{J}_a :=\lim_{\ep\rightarrow 0}J_a^\ep ~(a=0,1,2)$,
all operators are defined in the Hilbert space $\mathcal{H}$. The
specific
realization of $\mathcal{H}$ may be obtained by finding  eigenfunctions
of
the set of all commuting observables of the system. It is easy to verify
that for our system the commuting observables are $\hat{C}$ and
$\hat{J}_0$. The set of common eigenfunctions may be chosen to be
\begin{equation}
\phi_m (J,\beta) = \exp (im\beta),~~~~m\in \dZ ,
\end{equation}
which spans $ L^2(\dS^1,\mu)$ with $\mu (dx):= d\beta /2\pi$.

The Bargmann classification is characterized by the ranges of $q$ and
$m$. The range of our parameter $\kappa$ is $0< \kappa < \infty$, so it
corresponds to Bargmann's $1/4 < q < \infty$. Since $m\in\dZ$ in our
choice
of $\mathcal{H}$, the coherent state representation (in an asymptotic
sense)
we have found is almost everywhere identical  to the Bargmann's
continuous
integral case irreducible unitary representation called $C^0_q$,  with
$1/4 \leq q < \infty$, and this  corresponds  to the principal series
representation of $SO_0(1,2)$ group \cite{PJS}. The only difference is
that for massive particle we have $m_0 >0$, thus $\kappa = m_0 r_0 >0$,
so $q>1/4$. Our representation coincides with $C^0_q$ representation in
case taking the limit $\kappa\rightarrow 0$, i.e. $m_0\rightarrow 0$,
makes sense.
We discuss this issue in our next paper \cite{JPW} in relation with the
   work \cite{DBR}.

\section{Discussion}

As it is mentioned in section I, in the group theoretical
quantization method there is a problem with the ordering of
canonical operators appearing in the process of mapping classical
observables into the corresponding quantum operators. One does not
know how to solve this problem in the case in which observables are
higher than
first order polynomials in one half of the canonical coordinates
and related no-go theorems (see  \cite{MJG} for a recent review on
polynomial quantization problems). In the case of  particle dynamics
on hyperboloid we have managed to find such two canonical
variables that the basic observables (\ref{dyn2}) are linear
functions in one of them \cite{WP}, but our method cannot be
generalized to any observables.

On the other hand, there is no problem with the ordering of
operators in the coherent state quantization method presented
here. Indeed, the mapping (\ref{oper}) includes only classical
expression for observables. This is the main feature of the
method, since (\ref{oper}) may be applied to almost any  classical
observable, \emph{i.e.} smooth function of canonical variables,
and even to more exotic functions. The key ingredient of the
method is the definition of the coherent states, in particular the
construction of the set of orthonormal vectors (23) which span the
Hilbert space $\mathcal{K}_\ep$. We have solved this problem due
to the existence of a certain family of  coherent states
associated to the quantum motion of a particle on the unit circle.
It is clear that new physical problems call for new constructions
of $\mathcal{K}_\ep$.

As far as we know our paper presents the construction of the principal
series representation of $SO_0(1,2)$ group by making use of the
appropriate
coherent state quantization for the first time. Let us recall that the
representations of the
discrete series have already been obtained  within a coherent state
framework
\cite{JP1,AMP}.
It would be interesting to extend these results to include
the complementary series representations as well.

\appendix

\section{}

Application of the mapping (\ref{oper}) to the observables
$J_a~(a=0,1,2)$ reduces to the problem of calculating   quantities
like $A_\ep (J), ~A_\ep (e^{\pm i\beta})$ and $A_\ep (J e^{\pm
i\beta})$.

Making use of
\begin{equation}
\exp (\pm i\beta)=\cos\beta \pm i \sin\beta,~~~~~~
\int_{-\infty}^{+\infty}\exp{(-b t^2)} dt = \sqrt{\pi/b},~~~b>0,
\end{equation}
and Eq.(\ref{oper}), one can easily obtain the following
expressions
\begin{equation}
A_\ep (J)= \sum_{m=-\infty}^{+\infty}m|m><m|,~~~~~~
A_\ep (e^{i\beta})= e^{-\ep /4}\sum_{m=-\infty}^{+\infty}|m+1><m|,
\end{equation}
\begin{equation}
A_\ep (Je^{i\beta})= e^{-\ep /4}\sum_{m=-\infty}^{+\infty}(m+1)|m+1><m|.
\end{equation}
The formulae for $A_\ep (e^{-i\beta})$ and $A_\ep (Je^{-i\beta})$ are
obtained by taking the complex conjugate of
$A_\ep (e^{i\beta})$ and $A_\ep (Je^{i\beta})$, respectively.

\section{}

To show that the operators $\hat{J}_a ~~(a=0,1,2)$ are essentially
self-adjoint in $\mathcal{H}$, we make use of the Theorem 5 of the
Nelson
method \cite{EN}.

The Nelson operator, $\Delta$, for $sl(2,\dR)$ algebra reads
\begin{equation}
\Delta :=\hat{J}^2_0 + \hat{J}^2_1 + \hat{J}^2_2 =
\sum_{m=-\infty}^{+\infty}(2m^2 + \kappa^2 +1/4) |m><m|=\kappa^2 +1/4 +
2\sum_{m=-\infty}^{+\infty}m^2 |m><m|.
\end{equation}
It is clear that $\Delta$ is unbounded and symmetric in $\mathcal{H}$.
It is also essentially self-adjoint because it is a real diagonal
operator.

Since the operator $\Delta$ is essentially self-adjoint, it results from
the first part of Nelson's theorem that  our representation of
$sl(2,\dR)$
algebra is integrable to an unique unitary representation $U$ of the
universal covering group $\widetilde{SL(2,\dR)}$.

The second part of Nelson's theorem states that the application of the
Stone theorem \cite{RS} to the one-parameter unitary subgroups of $U$
generated by $J_a~(a=0,1,2)$ gives the essentially self-adjoint
operators
$\hat{J}_a ~(a=0,1,2)$.

\begin{acknowledgments}
The authors would like to thank J. Renaud for fruitful suggestions
and K. Kowalski for useful correspondence.

\end{acknowledgments}

\end{document}